\documentclass[prl,twocolumn,superscriptaddress]{revtex4}

\usepackage{amsmath}
\usepackage{amssymb}
\usepackage{xspace}
\usepackage{graphicx}
\usepackage{grffile}
\usepackage{nicefrac}
\usepackage{color}

\graphicspath{{figs/}}

\newcommand{\CoO}{CoO$_2$\xspace}
\newcommand{\NaxCoO}[1]{Na$_{#1}$CoO$_2$\xspace}

\newcommand{\Coa}{Co$^{3+}$\xspace}
\newcommand{\Cob}{Co$^{3.44+}$\xspace}
\newcommand{\Kagome}{kagom\'{e}\xspace}

\renewcommand{\Im}{\operatorname{\mathrm{Im}}}

\def\Ham{\hat{\mathcal{H}}}

\def\al{\alpha}
\def\bet{\beta}
\def\sig{\sigma}
\def\om{\omega}

\def\im{{i}}

\newcommand{\sd}{\downarrow}
\newcommand{\su}{\uparrow}

\def\kvec{\mathbf{k}}

\def\kv{\kvec}

\def\Kv{\mathbf{K}}

\newcommand{\ket}[1]{\left.\left|{#1}\right\rangle\right.}
\newcommand{\braket}[2]{\left\langle{#1}\middle| {#2}\right\rangle}

\newcommand{\copl}[2]{\hat{c}_{#1}^{#2}}
\newcommand{\cdagl}[2]{\hat{c}^{#2\dagger}_{#1}}

\newcommand{\nop}{\hat{n}}

\newcommand{\eps}{\varepsilon}

\begin{document}

\title{Strong correlations enhanced by charge-ordering in highly doped cobaltates}

\author{Oleg E. Peil}
\affiliation{I. Institut f{\"u}r Theoretische Physik,
Universit{\"a}t Hamburg, D-20355 Hamburg, Germany}
\author{Antoine Georges}
\affiliation{Centre de Physique Th\'eorique, \'Ecole Polytechnique, CNRS, 91128 Palaiseau Cedex, France}
\affiliation{Coll\`ege de France, 11 place Marcelin Berthelot, 75005 Paris, France}
\affiliation{DPMC, Universit\'e de Gen\`eve, 24 Quai Ernest Ansermet, 1211 Gen\`eve 4, Suisse}
\author{Frank Lechermann}
\affiliation{I. Institut f{\"u}r Theoretische Physik, Universit{\"a}t Hamburg, D-20355 Hamburg, Germany}

\begin{abstract}
We present an explanation for the puzzling spectral and transport properties 
of layered cobaltates close to the band-insulator limit, which relies on the 
key effect of charge ordering. 
Blocking a significant fraction of the lattice sites deeply modifies the electronic 
structure in a way that is shown to be quantitatively consistent with 
photoemission experiments. It also makes the system highly sensitive to interactions 
(especially to intersite ones), hence accounting for the strong correlations effects  
observed in this regime, such as the high effective mass and quasiparticle scattering rate.
These conclusions are supported by a theoretical study of an extended Hubbard model 
with a realistic band structure on an effective \Kagome lattice. 
\end{abstract}

\pacs{}

\maketitle
%
%
%
%
%

The layered cobaltate metals are famous for their remarkable electronic 
properties \cite{Foo_2004}, ranging from a large thermoelectric 
response \cite{Terasaki_1997}
charge-ordering \cite{Huang_2004, Alloul_2009} and puzzling magnetic behavior, 
to superconductivity when intercalated with water \cite{Takada_2003}.
%
Because of the universality of the physical properties  
throughout the cobaltate family, it is believed that most of these  
properties have their roots in the 
electronic structure of the \CoO layers. Those are formed of 
edge-sharing octahedra, with the Co ions forming a triangular 
lattice. Either alkali(-earth) metals (Na, Li, Ca etc., as in Na$_x$CoO$_2$) or 
more complex building blocks (e.g. rocksalt BiO or SrO planes
in misfit cobaltates \cite{Leligny_1999})
separate the layers and serve as electron donors. 
Varying the composition of the intercalated compounds
controls the doping $x \in [0,1]$, resulting in a 
nominal Co$^{(4-x)+}$ ($3d^{5+x}$) valence in the low-spin configuration
within the $t_{2g}$ manifold. The latter is split
by the local trigonal symmetry into one $a_{1g}$ and two
$e'_{g}$ states. Strong Coulomb interactions 
have been documented~\cite{Hasan_2004}
to occur within this orbital subspace.

Remarkably, most of the signatures of strong correlations in cobaltates, 
such as specific-heat enhancement \cite{Schulze_2008}, band narrowing
\cite{Hasan_2004, Nicolaou_2010, Nicolaou_2010b, Geck_2007} or 
strong electron-electron scattering \cite{Li_2004} 
are observed in the high-doping region 
$x \in [1/2,1]$ close to the band-insulating limit $(x=1)$. 
Besides intriguing magnetic behavior,
charge disproportionation is a prominent feature in this
part of the phase diagram. 
In \NaxCoO{3/4} clear evidences of separation of Co-ions into nonmagnetic \Coa-ions
and magnetic ions with a mixed valence are present \cite{Mukhamedshin_2005,Julien_2008}.
Importantly, this charge segregation is not restricted to 
selected dopings, but the concentration of \Coa-ions
is experimentally verified to steadily increase with doping 
starting from $x=1/2$ \cite{Lang_2008}.
Recently, the nuclear-magnetic-resonance (NMR)
experiments of Alloul {\sl et al.} \cite{Alloul_2009} revealed for $x=2/3$ a
charge-ordering pattern of cobalt ions with valency close to 
\Coa , with the remaining \Cob ions forming an effective \Kagome lattice (EKL).
The rather complex three-dimensional unit cell has common features revealed 
within a previous theoretical electronic-structure study~\cite{Hinuma_2008}. 
In general, multiple sodium-ordered phases are found in highly-doped sodium 
cobaltates~\cite{Roger_2007, Shu_2007, Hinuma_2008}.

In this Letter, we demonstrate that many puzzling physical properties of cobaltates at 
high doping may be understood within a picture of correlated itinerant electrons 
moving in a charge-ordered background. 
Charge-disproportionation patterns of \Coa-ions 
with well-localized electrons are formed, restricting the remaining 
carrier dynamics to an effective lattice with an open structure. 
Such a system is shown to be highly sensitive to correlations, 
both because of a prominent van-Hove singularity close to $x=0.7$ 
(owing to higher-order hopping terms) 
and due to the strong effect of intersite 
Coulomb interactions in line with the high susceptibility to 
charge-density wave (CDW) formation near $x=3/4$. 

To substantiate this view we consider a particular realization of charge ordering,
namely the EKL detected at $x=2/3$ for Na$_x$CoO$_2$ up to 
room temperature~\cite{Alloul_2009}. 
Whereas the charge-ordering at $x=1/2$ seems
unique to that stoichiometry, $x=2/3$ also marks the onset of 
Curie-Weiss behaviour and enhanced spin fluctuations, leading eventually to in-plane
ferromagnetic order for $3/4<x<0.9$. Furthermore, it is a commensurate doping
for CDW instabilities on the triangular 
lattice (as studied e.g. in Ref.~\cite{Motrunich_2004} within the Gutzwiller approach). 
Importantly, a recent calculation of the charge susceptibility in sodium cobaltate 
for the Hubbard model shows tendencies to the \Kagome instability at large $x$ even when 
disregarding the specific sodium distribution~\cite{Boehnke_2010}. 
Note however, that disordered donor configurations can by themselves enhance the 
correlation effects in the \CoO layers \cite{Marianetti_2007}, also magneto-polaronic 
excitations were suggested to play a role in this context~\cite{Khaliullin_2008}.
%
%
%
\begin{figure*}[ht]
\begin{center}
\includegraphics[width=\linewidth]{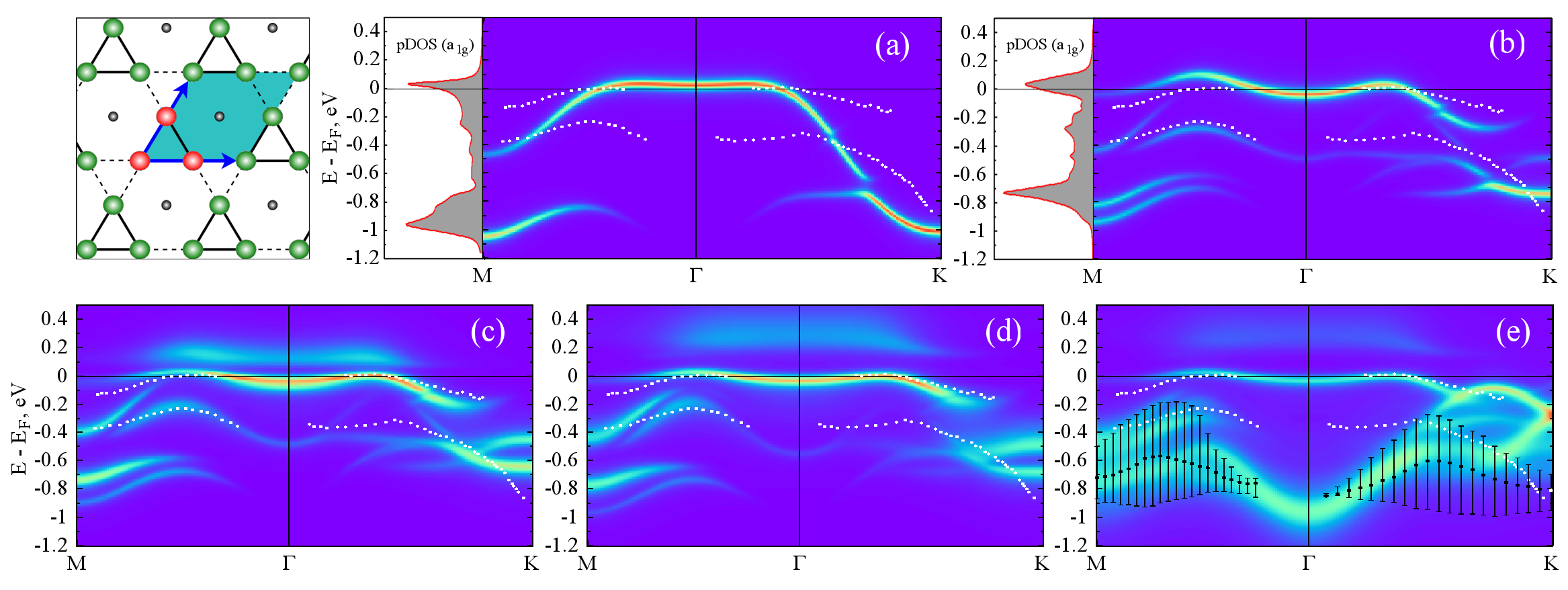}
\end{center}
\caption{(color online) Top left panel: Sketch of the EKL, with
blocked sites depicted by small spheres and the shaded region representing the unit cell. 
Panels (a-e): Unfolded spectral functions $A({\bf k},\omega)$ 
(according to Eq.~\eqref{eq:unfold})
projected onto $a_{1g}$ orbitals (a-d) and onto both $a_{1g}$ and the two $e'_{g}$ orbitals 
(e) for \NaxCoO{2/3} ($T=115$K). Top panels (a-b): results for
($U=V=0$) on the triangular lattice (a), and on the \Kagome lattice (b),
with the side-panels displaying the corresponding $k$-integrated spectral function.
Bottom panels (c-e): results obtained within CDMFT for
$U=5$ eV, $V=0$ (c) and for $U=5$ eV, $V=0.3$ eV (d), (e). 
The white dotted lines in (a-d) represent the QP peak and HE hump extracted from 
the ARPES data in Refs.~\onlinecite{Nicolaou_2010, Nicolaou_2010b}.
In (e) the black dotted lines correspond to the locations of the 
energy-distribution-curve (EDC) maxima of the spectral intensity seen at high
energies in the bare ARPES data of Refs.\onlinecite{Nicolaou_2010, Nicolaou_2010b} 
(with black bars showing the approximate width of the feature).
The quasiparticle and HE hump dotted lines in (e) are the same as in the other panels.}
\label{fig:pes}
\end{figure*}
%
%
%
%
%
In what follows, we address the spectral and transport properties of 
cobaltates stemming from a charge-order-induced EKL (see Fig.~\ref{fig:pes}), 
where the sodium and oxygen degrees of freedom
are integrated out. The EKL doping $y$ is related to the physical 
doping $x$ on the triangular lattice by $y = (4x - 1)/3$. 
The full $t_{2g}$ manifold is included in the modelling, however 
explicit two-particle interactions are only treated for $a_{1g}$, 
since it is known that the $e'_{g}$ bands lie below the Fermi 
level $\varepsilon_{\rm F}$ for the high-doping regime. 
Our realistic model on the EKL is defined by an 
extended Hubbard Hamiltonian $\Ham= \hat{H_{0}} + \hat{H}_{\rm int}$ with
\begin{eqnarray}
\hat{H_{0}}&=& - \sum_{\sig \langle ij \rangle \al \bet} t^{\al\bet}_{ij}
\cdagl{\sig i}{\al} \copl{\sig j}{\bet}
+\Delta \sum_{\sig i} \nop^{1}_{\sig i}\,\,, \label{eq:ham_0} \\
\hat{H}_{\rm int}&=&
U \sum_{i} \nop^{1}_{\su i} \nop^{1}_{\sd i} + V \sum_{\langle ij \rangle}
\nop^{1}_{i} \nop^{1}_{j}\,\,,
\label{eq:ham_int}
\end{eqnarray}
where the hopping matrix $t^{\al\bet}_{ij}$ is parametrized up to the
third nearest-neighbor (NN) from the band structure of \NaxCoO{x} 
within the local-density approximation. Orbital indices are
$\al = 1$ for $a_{1g}$ and $\al = 2,3$ for $e'_{g1}$, $e'_{2g}$. 
The parameter $\Delta$ fixes the $a_{1g}$-$e'_{g}$ renormalized crystal-field splitting. 
We choose $U = 5$ eV in line with experimental and theoretical 
determinations \cite{Hasan_2004, Kroll_2006} and varied
the NN intersite Coulomb interaction $V$ up to $0.3$ eV. 
Decreasing $U$ down to 3.5 eV provided no significant variation of the results.

The band structure of the noninteracting realistic model consists of nine
bands inside the Brillouin zone (BZ) of the EKL.
To connect to the results of angle-resolved photoemission
(ARPES) experiments we represent the computed
$k$-resolved spectral function in the BZ of the underlying triangular lattice. 
This can be achieved, following the method of Ref.~\onlinecite{Ku_2010}, 
by "unfolding" the nine bands in the supercell \Kagome-lattice BZ (SBZ) to the three
bands in the normal BZ (NBZ) of the triangular lattice. The states 
$\ket{\Kv \al}$ defined within the SBZ are projected onto
the states $\ket{\kv \al}$ defined in the NBZ, so that the spectral function projected 
on orbital $\al$ reads:
\begin{align}
A_{\alpha}(\kv,\om)= -\frac{1}{\pi} \sum_{\Kv l l'}\hspace*{-0.1cm} 
\braket{\kv\al}{\Kv l} 
\Im G_{l l'}(\Kv,\om)\braket{\Kv l'}{\kv\al},
\label{eq:unfold}
\end{align}
where $\Kv$ $\in$ SBZ and $\kv$ $\in$ NBZ. The
lattice Green's function,
$
G_{l l'}(\Kv, \om ) = \left[ 
(\om + \mu) \delta_{ll'} - H_{0,l l'}(\Kv) -
\Sigma_{l l'}(\Kv, \om ) \right]^{-1}
$
is written in the basis $\ket{l} \equiv \ket{i,\al}$,
where $i$ runs over the three atoms of the \Kagome-lattice unit cell.

%
%
%

To examine the spectral properties, we project $A(\Kv,\om)$ onto the 
$a_{1g}$ and $e'_{g}$ orbitals and unfold the EKL bands into the NBZ.
The obtained spectral functions are compared to the ARPES data 
of Nicolaou {\it et al.} for $x = 0.7$ \cite{Nicolaou_2010, Nicolaou_2010b}. 
Although those data were obtained for misfit cobaltates (BiO and BaO planes), 
at present stage the spectra obtained for sodium cobaltates are 
qualitatively very similar (see e.g.~\cite{Qian_06}, \cite{Geck_2007}).
The analysis of the polarization-dependence of the ARPES spectra performed by the 
authors of Ref.~\cite{Nicolaou_2010, Nicolaou_2010b} led them to identify
two $a_{1g}$-like features: a low-energy quasiparticle (QP) peak close to 
$\varepsilon_{\rm F}$,  and a high-energy (HE) hump around -0.4 eV, 
both shown as dotted lines in Fig.~\ref{fig:pes}.

Already for $U=V=0$ the theoretical spectral function
of the EKL is found to be in much better agreement with
experiment than that of the original triangular structure without
charge order, as shown in Fig.~\ref{fig:pes}a, b. 
The HE hump identified experimentally, and completely absent in the triangular-lattice model, 
matches quite well with the HE feature around -0.4 eV in the EKL model. 
This HE feature originates here from 
$a_{1g}$-$e'_{g}$ hybridizations (cf. Fig.~\ref{fig:pes}e
where the full $t_{2g}$ projection is shown)
induced by the loss of the original full triangular symmetry. 
It is thus essential to keep the $e'_{g}$ states in the model to reproduce the spectral 
structure in this energy range around the $\Gamma$-point.
In contrast, in Ref.~\onlinecite{Nicolaou_2010, Nicolaou_2010b}
the HE feature was interpreted as an incoherent excitation
with $a_{1g}$ character of unknown type.
However, it should be noted that the full local triangular symmetry with equivalent Co sites  
together with a compensation between the spectra of the two $e^\prime_g$ orbitals
were assumed in the analysis of Ref.~\onlinecite{Nicolaou_2010, Nicolaou_2010b} 
in order to extract an $a_{1g}$ signal. 
Since triangular symmetry is broken in the charge-ordered phase, 
refinements in the polarization-dependent 
analysis of the ARPES data may be required. 

The full interacting model is solved by dynamical mean-field theory 
(DMFT) and its cellular-cluster extension (CDMFT) \cite{Lichtenstein_2000, Biroli_2004}. 
A continuous-time quantum Monte Carlo solver
within the hybridization-expansion approach \cite{Gull_2011} is used, in the 
implementation by Parcollet and Ferrero \cite{Ferrero_09}.
The cluster impurity model is defined in terms of a natural
basis of molecular orbitals via 
$
\ket{d_{1}} = (-2 \ket{1} + \ket{2} + \ket{3}) / \sqrt{6},
\ket{d_{2}} = (\ket{2} - \ket{3}) / \sqrt{2},
\ket{u} = (\ket{1} + \ket{2} + \ket{3}) / \sqrt{3},
$
where $\ket{i}$, $i = 1,2,3$ denote states corresponding
to the basis atoms of the \Kagome-lattice unit cell.
In this representation the impurity self-energy matrix is diagonal
with elements
$(\Sigma_{d}, \Sigma_{d}, \Sigma_{u}) = (\Sigma_{0} - \Sigma_{1}, 
\Sigma_{0} - \Sigma_{1}, \Sigma_{0} + 2\Sigma_{1})$,
where $\Sigma_{0}$, $\Sigma_{1}$ are the on-site and
intersite self-energies, respectively. 

\begin{figure}
\begin{center}
\includegraphics[width=1.0\linewidth]{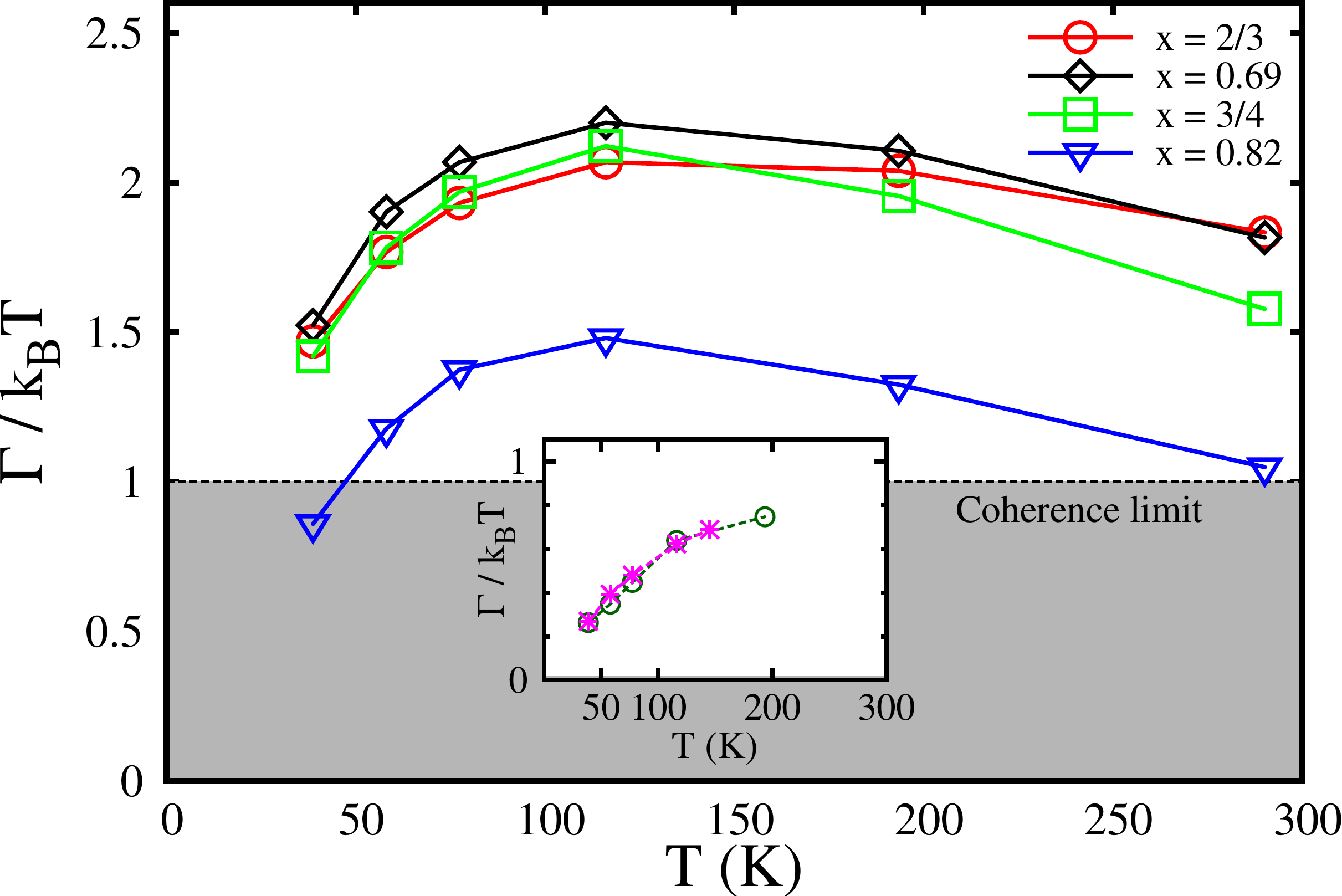}
\end{center}
\caption{(color online) Temperature-dependence of 
$\Gamma/k_{\mathrm{B}}T$, with $\Gamma$ the inverse QP lifetime 
for the \Kagome cobaltate model with on-site Coulomb 
interaction $U$ ($V = 0$) for different $x$. 
Inset: $\Gamma/k_{\mathrm{B}}T$ for the triangular cobaltate model including distant hoppings
at $x = 3/4$ (green, circles) and for the \Kagome lattice with 
only NN hoppings at $x = 2/3$ (magenta, stars).}
\label{fig:scat_rate}
\end{figure}

\begin{figure}
\begin{center}
\includegraphics[width=1.0\linewidth]{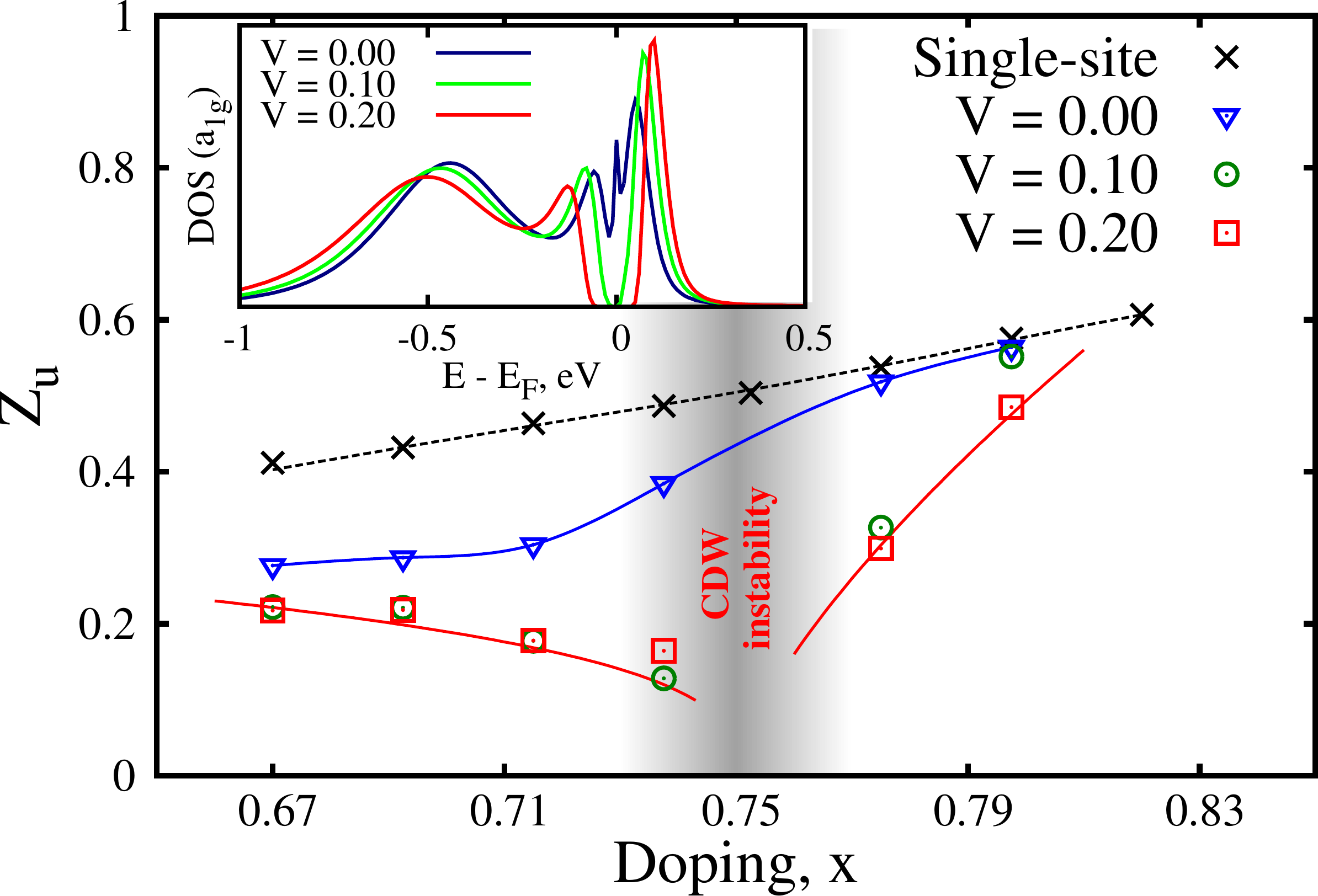}
\end{center}
\caption{(color online) Doping-dependence of $Z$ for the
symmetric molecular orbital $\ket{u}$ of
the \Kagome cobaltate model at $T = 39K$. 
Crosses: single-site calculations, thus $Z_{u} = Z_{d} = Z$;
triangles, circles, boxes: cluster calculations.
All lines are just guides to the eye. Inset: 
the local spectral function projected onto the $a_{1g}$ 
orbital with $V$ at $x = 3/4$. }
\label{fig:z_vs_dop}
\end{figure}

%
%
%
In terms of spectral properties, the on-site $U$-term alone ($V = 0$) already yields  
a significant renormalization of the QP band: its bandwidth is 
reduced by a factor $Z\simeq 0.4$ 
(compare panels (b) and (c) in Fig.\ref{fig:pes}). Spectral-weight transfer takes place 
to an incoherent upper Hubbard band above $\varepsilon_{\rm F}$ (Fig.~\ref{fig:pes}c). 
But the strongest effect of $U$ actually manifests itself in a high value of 
the inverse QP lifetime $\Gamma\equiv -Z\Im \Sigma(\im 0^{+})$.  
Figure~\ref{fig:scat_rate} displays the dimensionless ratio 
$\Gamma/k_{\mathrm{B}}T$ vs. $T$ evaluated in DMFT. 
Remarkably, this ratio is significantly larger than unity for  
almost all temperatures and doping studied.
Thus, long-lived coherent QPs only exist 
at very low $T$, due to electron-electron scattering. 
Extrapolation of our numerical results suggest that the coherent regime 
($\Gamma/k_{\mathrm{B}}T < 1$) would be reached 
only for $T\lesssim 20$K at $x=2/3$, and the true Fermi-liquid behaviour 
($\Gamma/k_{\mathrm{B}}T \propto T$) at an even lower temperature scale. 
This is qualitatively consistent with the large values of the resistivity 
reported in this doping regime (e.g. $\rho\sim 300\mu\Omega$cm for 
Na$_{0.71}$CoO$_2$ at $T=100$K~\cite{Foo_2004}), as well as with the 
huge scattering rate at low $T$~\cite{Li_2004} and the low value of the saturation
temperature of the susceptibility \cite{Mukhamedshin_2004}. 
The value of $\Gamma$ is largest for $x\simeq 0.69$, which is approximately 
the doping at which the van-Hove
singularity lies at the Fermi level (see side-panel of Fig.~\ref{fig:pes}b). 
Notably, only the EKL model with realistic hoppings shows this characteristics, 
while sole NN hoppings
(having no flat dispersions near $\eps_{\rm F}$ at large $x$ for $t<0$)
lead to a low scattering rate
(inset of Fig.~\ref{fig:scat_rate}). 
%
%
%
%
%
%
%
%
The behavior of the system is strongly modified 
when the intersite interaction, $V > 0$,
is taken into account.
Figure~\ref{fig:z_vs_dop} displays the QP weight $Z_u$ 
from CDMFT for the `symmetric' molecular orbital 
of the EKL as a function of $x$ for increasing $V$. 
For $x=2/3$, $V$ has only moderate effect, although it does reduce 
$Z$ by a factor of two, bringing it close to the experimentally observed 
mass enhancement ($m^*/m_{\mathrm{LDA}}\simeq 4$). Some improvement 
in the comparison to ARPES at this doping is also found (Fig.~\ref{fig:pes}d,e). 
In contrast, close to $x=3/4$, the system is highly sensitive to $V$. Indeed, this 
doping level corresponds to the commensurate doping $y=2/3$ on the EKL,
at which a nonzero $V$ can induce a CDW instability towards an insulating
state. We find this to happen at a rather small value of $V_c=0.1$~eV
(inset of Fig.~\ref{fig:z_vs_dop}). 
This renders it possible that a very strongly correlated metallic state 
(with a small QP weight renormalized by charge fluctuations) found close to $x=3/4$, is 
best described as a weakly doped CDW insulator. 
It is also worth mentioning that $x=3/4$ is the doping for the onset 
of magnetic ordering, and that transport anomalies were observed at this doping
\cite{Motohashi_2003}. 

Note that previous works already showed the importance of nonlocal 
correlations~\cite{Piefke_2010,Li_2011}.
From a `molecular orbitals' viewpoint, the CDW transition 
is driven by orbital polarization: the intersite interaction drives the symmetric orbital 
$|u\rangle$ towards half-filling (hence susceptible to a MIT), while the two 
asymmetric orbitals $|d_{1,2}\rangle$ are completely filled 
(with in total $5=3(1+y)$ electrons in each EKL unit cell). 
%
%
%
%

In conclusion, we have shown that several of the intriguing physical properties 
of layered cobaltates at large $x$ can be explained within a 
picture of electrons moving in a charge-ordered background. 
In particular, good quantitative agreement with ARPES spectra has been obtained. 
Although we have considered a specific realization for the charge-ordered phase
in the form of a \Kagome structure, we expect similar effects for
other ordering patterns in which a large fraction
of Co ions localize charge and thus become blocked for electron hopping processes. 
This blocking makes the system highly sensitive to correlation effects, in 
particular to intersite Coulomb repulsion. The proximity of a van Hove singularity 
has also been shown to play a key role. 
Interesting issues to be addressed in future 
work are whether such charge blocking can account 
for the large thermopower observed in this doping regime~\cite{Haule_2009}
and whether it can provide a mechanism for the apparent violation of the
Luttinger sum rule \cite{Nicolaou_2010}.

\begin{acknowledgments}
We are grateful to V. Brouet and A.Nicolaou for extensive discussions of their ARPES data, 
and also acknowledge useful discussions with M.Aichhorn, H.Alloul, L. Boehnke, 
A.I. Lichtenstein and J.Mravlje. Calculations were performed at the North-German 
Supercomputing Alliance (HLRN). This research was supported in part by the National 
Science Foundation under Grant No. NSF PHY05-51164 as well as the SPP1386 and the 
FOR1346 project of the DFG, and the Partner University Fund.
\end{acknowledgments}


\end{document}